\documentclass[11pt]{article}
\usepackage{amssymb,latexsym, amsmath}
\input epsf
\usepackage{amscd}

\makeatletter \@addtoreset{figure}{section}
\def\thefigure{\thesection.\@arabic\c@figure}
\def\fps@figure{h, t}
\@addtoreset{table}{bsection}
\def\thetable{\thesection.\@arabic\c@table}
\def\fps@table{h, t}
\@addtoreset{equation}{section}

\newif\ifamsfonts
\amsfontstrue \ifamsfonts
\font\twlbbb=msbm10 scaled\magstep1 \font\egtbbb=msbm8
\font\sixbbb=msbm6
\newfam\mathbbfam
\textfont\mathbbfam=\twlbbb \scriptfont\mathbbfam=\egtbbb
\scriptscriptfont\mathbbfam=\sixbbb

\makeatother

\newtheorem{theorem}{Theorem}[section]
\newtheorem{proposition}[theorem]{Proposition}

\newfont{\tenbi}{cmbxti10}

\begin{document}

\title{Integrable flows and Backlund transformations on extended Stiefel varieties 
with application to the Euler top on the Lie group $SO(3)$. 
\footnote{AMS Subject Classification 58F07, 70H99, 76B15} }

\author{Yuri N. Fedorov \\
Department of Mathematics and Mechanics 
\\ Moscow Lomonosov University, Moscow, 119 899, Russia \\ 
e-mail: fedorov@mech.math.msu.su \\ 
and \\ 
 Department de Matem\`atica I, \\ 
Universitat Politecnica de Catalunya, \\
Barcelona, E-08028 Spain \\
e-mail: Yuri.Fedorov@upc.es  }
\maketitle

{\it Short title}: Flows and Backlund transformations on extended Stiefel varieties 

\begin{abstract} 
We show that the $m$-dimensional Euler--Manakov top on $so^*(m)$ can be represented
as a Poisson reduction of an integrable Hamiltonian system on a symplectic extended 
Stiefel variety $\bar{\cal V}(k,m)$, 
and present its Lax representation with a rational parameter.

We also describe an integrable two-valued symplectic map  $\cal B$ on the 4-dimensional variety
${\cal V}(2,3)$. 
The map admits two different reductions, namely, to the Lie group $SO(3)$ and to the coalgebra $so^*(3)$.

The first reduction provides a discretization of the motion of the classical Euler top in space
and has a transparent geometric interpretation, which can be regarded as a discrete version of the
celebrated Poinsot model of motion and which inherits some properties of another discrete system, 
the elliptic billiard.  

The reduction of  $\cal B$ to $so^*(3)$ gives a new explicit discretization  of the Euler top
in the angular momentum space, which preserves first integrals of the continuous system. 
\end{abstract}

\section{Introduction}
In most publications the integrable $m$-dimensional Euler top is represented as a flow on  
the cotangent bundle $T^* SO(m)$ or on the coalgebra $so^*(m)$.

Recently, an alternative description of this problem as a system on a 
symplectic subvariety of the group product 
$SO(m) \times SO(m)$ was  proposed in \cite{Bl_Cr, BCMR}. 
 
A first discretization of the free $m$-dimensional top on $T^* SO(m)$ was 
constructed in \cite{Ves, MosVes}
by the method of factorization of matrix polynomials. 
This discretization is represented by a second order Lagrangian correspondence,
which does not explicitly involve a time step, it is determined by initial data 
(a choice of two subsequent points on $SO(m)$). 

On the other hand, in \cite{S} (see also \cite{SB}) 
Suris introduced a concept of an integrable discretisation of a 
finite-dimensional Hamiltonian system as a one parameter family of integrable Poisson maps 
parameterized by a time step  $\epsilon$, which 
differ from the identity map by $O(\epsilon)$, and whose Poisson structure and the  integrals of motion 
differ at most by  $O(\epsilon)$ from those of the continuous-time system. 

In the special case where the discretization preserves exactly both the Poisson structure 
and the integrals of motion, one speak of an ``exact discretization'': 
one has a family of B\"acklund transformations, which map solutions into solutions 
and are interpolated by a hamiltonian flow generated by some function of the integrals of 
motion of the continuous system. 

A class of implicitly defined Poisson maps $so^*(3) \to so^*(3)$ discretizing the classical Euler top 
in the space of the angular momentum  was indicated in \cite{Bob_Lorb}. 
The maps preserve the energy and momentum integrals  of  the continuous problem and contain explicitly 
a time step parameter.
It was shown that such a map preserves the standard Lie--Poisson structure on $so^*(3)$  
if and only if its restriction onto complex invariant manifolds, open subsets of elliptic curves,
is a shift, which is constant on each curve. 

Recently, another integrable discretization of the top on $so^*(3)$, which is {\it explicit}, but does not preserve 
the integrals of the continuous problem was found in \cite{Jap} by applying the Hirota method.
\medskip

\paragraph{Contents of the paper.} Our aim is twofold. First, in Section 2,
we propose yet another description of the continuous $n$-dimensional Euler--Manakov top as a reduction of
a Hamiltonian system on so called extended Stiefel variety $\bar{\cal V}(k,m)$,  a symplectic submanifold 
of dimension $km-k^2/2$ in ${\mathbb R}^{km}$, where $2\le k\le n$ is an even integer. 
We present a Lax representation of this system with a rational parameter, which, in a sense described
below, is dual to Manakov's Lax pair found in \cite{Man_so_n}.
 
The system possesses $k/2$ commuting symmetry fields  ${\cal R}_l$ generated by Hamiltonians $H_l$. 
Its  Marsden--Weinstein reduction with respect to the action of  the fields gives rise to 
a Hamiltonian system on a rank $k$ coadjoint orbit ${\cal S}_h^{(k)}$ in the coalgebra $so^*(m)$, 
whereas the original Poisson structure in $\bar{\cal V}(k,m)$ is a pull-back of 
 the standard Lie--Poisson structure of  $so^*(m)$ restricted onto the orbit. 
The reduced Hamiltonian system coincides with the  Euler--Manakov system on
${\cal S}_h^{(k)}$. In case of the maximal rank $k$, the level variety $\{H_l=c_l \}\subset \bar{\cal V}(k,m)$ 
is the group $SO(m)$, and the restriction of the original system onto the group yields a flow describing
the motion of the $n$-dimensional top in space. 

Second, in Section 3, we present an intertwining relation (discrete Lax pair) generating 
a explicit $\lambda^*$-depended family of two-valued complex B\"acklund 
transformations  $\cal B_\lambda^*$ of the variety ${\cal V}(2,3)$, 
which preserve the above Poisson structure and the first integrals of the continuous 
Hamiltonian system (formula (\ref{4.10})).  

The restricton of  $\cal B_\lambda^*$ onto the group $SO(3)$ 
provides a discretization of the motion of the classical Euler top in space and
has a transparent geometric interpretation, which, in turn, 
can be regarded as a discrete version of the
celebrated Poinsot model of motion and which inherits some properties of another discrete integrable system, 
the elliptic billiard (Figure 1).  

On the other hand, the reduction of  $\cal B_\lambda^*$ onto the coalgebra $so^*(3)$ gives a new explicit discretization  
of the classical Euler top, which also preserves its first integrals (formula (\ref{diff})). 

Like the Moser--Veselov correspondence,
the both discretizations do not explicitly involve a time step and their continuous limits depend on  
the parameter $\lambda^*$.

\section{Hamiltonian Systems on Extended Stiefel Varieties 
and Rank $k$ Solutions of Frahm--Manakov top}

Recall that the free motion of an $m$-dimensional rigid body is described by 
the Euler--Frahm equations (\cite{FFrahm}) 
\begin{equation} 
\dot M=[M,\Omega ],  
\label{2.1}
\end{equation}
where $\Omega \in so(m)$ is the angular velocity,
$M\in so^*(m)$ the angular momentum of the body in the moving frame. 
Following \cite{MiFo, R}, these equations  are Hamiltonian 
with respect to  the degenerate Lie--Poisson bracket on $so^*(m)$ 
\begin{equation} \label{poinsot}
\{  M_{ij},  M_{kl} \}_{so(m)} = \delta_{il} M_{jk} - \delta_{il} M_{kj}
+\delta_{kj} M_{jl} - \delta_{ik} M_{jl}  
\end{equation}
and $\Omega_{ij}=\partial H(M)/\partial M_{ij}$. 

The restriction of $\{\cdot, \cdot  \}_{so(m)}$ onto orbits of coadjoint action of
$SO(m)$  in $so^*(m)$ is nondegenerate.  A generic orbit ${\cal S}_h$ parameterized by 
$[m/2]$ independent Casimir functions of the bracket is thus a symplectic variety of dimension
$m(m-1)/2-[m/2]$. 


Equations (\ref{2.1}) are known to
be integrable provided $M$ and $\Omega$ are related as
$[M, a]=[\Omega, b]$, where $a,b$ are constant commuting matrices, 
and all the eigenvalues of $a$ and $b$ are distinct.
The integrability follows from the  Lax representation
with a rational spectral parameter found by Manakov in \cite{Man_so_n}, 
or from 
a hyperelliptic Lax pair indicated in \cite{Fe_AMS}. These Lax pairs provide a complete set
of integrals of motion, whose involutivity  can be proved by applying $r$-matrix theory.

For the concreteness, 
in the sequel we consider the case $a=\mbox{diag}(a_1,\dots,a_m)$, $b=a^2$.
Then $\Omega=AM+M A$, and equations (\ref{2.1}) take the form
\begin{equation}  
\label{FM_again}
\dot M=[M,aM+Ma]
\end{equation}

 Apart from this ``basic'' system, there exists a whole hierarchy of
``higher Manakov systems'',  which are defined by different relations between $\Omega$ and $M$,
and which commute with (\ref{FM_again}).
\medskip

Below we show that the restrictions of the Frahm--Manakov system on rank $k$ orbits of coadjoint representation 
of $SO(m)$ in  $so^*(m)$ are closely related to certain Hamiltonian 
dynamical systems on extended Stiefel varieties. 
Recall that the standard Stiefel variety ${\cal V}(k,m)$ is
the variety of ordered sets of $k$ orthogonal vectors in ${\mathbb R}^m$ (${\mathbb C}^m$) 
having fixed squares. It is a smooth variety of dimension $km-k(k+1)/2$ (see e.g., \cite{Sovr_Geom}).

Namely, as follows from (\ref{2.1}), the angular momentum in space is a constant matrix.
Hence, due to the Darboux theorem, in the case rank $M=k$ there exist $k$ mutually orthogonal and
{\it fixed in space\/} vectors $x^{(l)}, y^{(l)}\in {\mathbb R}^m$, $l=1,\dots,k/2$ such that
$|x^{(l)}|^2=|y^{(l)}|^2=h_l$ and the momentum $M$ can be represented in form
\begin{gather}
M=\sum_{l=1}^{k/2} x^{(l)}\wedge y^{(l)}\, , \quad \mbox{that is,} \quad
M= {\cal X}^T{\cal Y}-{\cal Y}^T{\cal X} \, , \label{r5.87} \\
{\cal X}^T=(x^{(1)}\, \cdots \, x^{(k/2)})\, , \quad {\cal Y}^T=(y^{(1)}\, \cdots \,
y^{(k/2)})\, . 
\nonumber
\end{gather}
Under the above conditions, 
the set of $k\times m$ matrices ${\cal Z}=( x^{(1)}\,y^{(1)}\,\cdots\, x^{(k/2)}\,y^{(k/2)})^T$
forms the {\it extended \/}  Stiefel variety $\bar{\cal V}(k,m)$.
In contrast to  the standard Stiefel variety, for each index $l$,
the absolute values  $|x^{(l)}|=|y^{(l)}|$ are not fixed. 
Thus, $\bar{\cal V}(k,m)$ is of dimension $km-k^2/2$, and the $k\times m$ components of $\cal Z$
play the role of excessive coordinates on it. 

Let 
$$
{\omega}=\mbox{tr}\, (d {\cal X} \wedge d {\cal Y}^T)
=\sum_{l=1}^k \sum_{i=1}^m \, d x^{(l)}_i \wedge d\,y^{(l)}_i\, 
$$
be the canonical symplectic structure on the space 
${\mathbb R}^{km}=\widetilde x^{(1)},y^{(1)},\cdots , x^{(k/2)},y^{(k/2)})$
and let $\bar\omega$ denote the restriction of 2-form $\omega$ onto 
$\bar{\cal V}(k,m)\subset{\mathbb R}^{km}$. The latter subvariety
is defined by conditions
\begin{gather} \label{constraints}
 \langle x^{(l)}, x^{(l)}\rangle -  \langle y^{(l)}, y^{(l)}\rangle =0, \quad  
\langle x^{(l)}, x^{(s)}\rangle 
= \langle y^{(l)}, y^{(s)}\rangle =0, \quad
{\cal Y}{\cal X}^T=0, \\
l,s=1,\dots, k/2,  \nonumber
\end{gather}
which consist of $k^2/2$ independent scalar equations $f_s(x,y)=0$. 
The matrix of standard Poisson brackets of the constraint functions $f_s$ in 
${\mathbb R}^{km}$ is nondegenerate. 
It follows that 2-form $\bar\omega$  is also nondegenerate
and the extended Stiefel variety is symplectic.

Since the vectors  are fixed in space, in the frame attached to the top they satisfy the Poisson
equations $\dot x^{(l)}=-\Omega x^{(l)}$, $\dot y^{(l)}=-\Omega y^{(l)}$,
$\Omega\in so(m)$, which imply
\begin{equation}
\dot{\cal X}={\cal X}\Omega\, , \quad \dot{\cal Y}={\cal Y}\Omega \, .
\label{r5.88}
\end{equation}
As above, we put $\Omega=a M+M a$, $a=\mbox{diag} (a_1,\dots,a_m)$
and define a dynamical system on $\bar{\cal V}(k,m)$, which is generated by
(\ref{r5.88}), (\ref{r5.87}):
\begin{equation}
\begin{aligned}
\dot{\cal X} &={\cal X}[ a({\cal X}^T {\cal Y}-{\cal Y}^T{\cal X})
+{\cal X}^T{\cal Y}a],  \\
\dot{\cal Y} &={\cal Y} [ a ({\cal X}^T{\cal Y}-{\cal Y}^T{\cal X})
-{\cal Y}^T{\cal X} a ] \, .  \end{aligned} \label{r588}
\end{equation}

\begin{theorem}\label{tr5.25}\begin{description}
\item{\rm 1)} Under the substitution $(\ref{r5.87})$
solutions of the system $(\ref{r588})$ give rank $k$ solutions of the Frahm--Manakov
system (\ref{FM_again}) on $so^*(m)$. 

\item{\rm 2)}
Up to the action of the discrete group generated by reflections
$({\cal X},{\cal Y})\to (-{\cal X},-{\cal Y})$, the system $(\ref{r588})$
is described by $k\times k$ Lax pair with rational parameter $\lambda$
\begin{gather}
\dot L(\lambda)=[\, L(\lambda), A(\lambda)\,]\, , \qquad L,A\in {\rm sp}(k/2), \quad
\lambda\in{\mathbb C} \, , \label{r5.89} \\
L=\begin{pmatrix}-{\cal X} (\lambda{\bf I}-a)^{-1} {\cal Y}^T &
-{\cal X} (\lambda{\bf I}-a)^{-1} {\cal X}^T \\
{\cal Y} (\lambda{\bf I}-a)^{-1} {\cal Y}^T &
{\cal Y} (\lambda{\bf I}-a)^{-1} {\cal X}^T \end{pmatrix}
\equiv \sum_{i=1}^m  \frac{ {\cal N}_i} {\lambda-a_i}\, , \label{dual1} \\
{\cal N}_i=\begin{pmatrix} {\bar x}_i {\bar y}_i^T & -{\bar x}_i {\bar x}_i^T \\
-{\bar y}_i {\bar y}_i^T & -{\bar y}_i {\bar x}_i^T \end{pmatrix}, \nonumber  \\
A=\begin{pmatrix}
{\cal X} (a+\lambda{\bf I}){\cal Y}^T  & {\cal X} (a+\lambda{\bf I}){\cal X}^T \\
-{\cal Y} (a+\lambda{\bf I}) {\cal Y}^T & -{\cal Y} (a+\lambda{\bf I}){\cal X}^T
\end{pmatrix}\,, \label{AA}
\end{gather}
where $\bar x_i=\left( x_i^{(1)},\dots,x_i^{(k/2)} \right)^T$
(respectively $\bar y_i=\left( x_i^{(1)},\dots,x_i^{(k/2)}\right)^T)$ is $i$-th column of $\cal X$
{\rm (}respectively of $\cal Y)$, and ${\bf I}$ is the unit $m\times m$ matrix.
\end{description}
\end{theorem}

\noindent{\it Proof.} The first statement follows directly from
the derivation of the system $(\ref{r588})$. Further, we calculate the derivative 
$\dot L(\lambda)$ by virtue of equations (\ref{r588}).
In view of matrix relations 
$(\lambda{\bf I}-a)^{-1} a=\lambda(\lambda{\bf I}-a)^{-1}-{\bf I}$ and 
${\cal Y}{\cal X}^T={\cal X}{\cal Y}^T=0$,  the derivative  coincides 
with the commutator in $(\ref{r5.89})$. $\boxed{}$
\medskip

\noindent{\bf Remark 2.1.} 
Notice that the entries of matrices 
$$
\Phi(\lambda)L(\lambda), \quad \Phi(\lambda)A(\lambda), \quad \mbox{where} \quad
\Phi(\lambda)=(\lambda-a_1)\cdots (\lambda-a_n)
$$  
are polynomials in $\lambda$, and, under the substitution (\ref{r5.87}),
the coefficients of the characteristic polynomial $| \Phi(\lambda) L(\lambda)-w{\bf I}|$
can be expressed in terms of $M_{ij}$ only as follows 
\begin{gather}
|w{\bf I}-L(\lambda)|=w^k +\sum_{l=2}^k w^{k-l} \Psi^{l-1}(\lambda)\, 
\widetilde {{\cal I}}_{l}(\lambda,M) \, , \qquad l=2,4,\dots,k, \nonumber \\
\widetilde {\cal I}_{l}(\lambda,M)=\sum^{m}_{I}
\frac{\Phi(\lambda) }{ (\lambda-a_{i_1})\cdots(\lambda-a_{i_l})}\,|M|^{I}_{I} \, ,
\label{cal_I}
\end{gather}
where $|M|^{I}_{I}$ are diagonal minors of order $l$ corresponding to
multi-indices $I=\{i_1,\ldots i_k\}\subset\{1,\dots,m\}$, $1\le i_1<\cdots<i_l\le m.$
Notice that the leading coefficients $H_{l,m-l}(M)=\sum^{m}_{I} \,|M|^{I}_{I}$ form
a complete set of Casimir functions on $so^*(m)$. 
\medskip

The $k\times k$ matrix $L(\lambda)$ in (\ref{dual1}) belongs to a wide class of Lax operators of the form 
$$
Y+\sum_{i=1}^n \frac{ G_i^T F_i}{\lambda-a_i},  
$$
where $Y\in gl(k)$ is a constant matrix and $G_i, F_i$ are $k_i\times k$ matrices.
Such Lax matrices can be regarded as images of moment maps to the loop algebra 
$\widetilde {gl}(k)$, and integrable systems generated by them 
have been studied in the series of papers \cite{AHP,AHH_CMP,AHH_Lett,Harnad1}
in connection with the duality to so called rank $k$ perturbations of
constant diagonal matrices of dimension $n\times n$ (following Moser \cite{Moser_Chern}). 
In particular, the  $k\times k$  Lax matrix (\ref{dual1}) 
is dual to the  $n\times n$ Lax matrix in the Manakov representation,
\begin{equation}
{\cal L} (\mu) = a +\frac 1\mu ({\cal X}^T{\cal Y}-{\cal Y}^T{\cal X} )\equiv 
  a +\frac 1\mu M,  \label{2.17} 
\end{equation}
in the sense that under the relation (\ref{r5.87})  
the  spectral curves $|L(\lambda)- \mu {\bf I}|=0$ and
$|{\cal L}(\mu)- w{\bf I}|=0$ are birationally equivalent and 
the parameter $\lambda$ plays the role of the eigenvalue parameter for (\ref{2.17}).
The characteristic polynomials of the dual Lax matrices are related by the 
Weinstein--Aronjan formula (see \cite{AHP}). 
\medskip

\noindent{\bf Remark 2.2.} The matrix $A(\lambda)$ in (\ref{AA}) can be represented in form
$$
A(\lambda)\, =[\, \lambda^{-m+2} \Phi(\lambda) L(\lambda) \, ]_+ + (a_1+\cdots +a_n) L_0 ,  \qquad
L_0=\begin{pmatrix}  0 &  {\cal X} {\cal X}^T  \\
                                    -{\cal Y} {\cal Y}^T &  0 \end{pmatrix}
$$
where $[\;\; ]_+$ denotes the polynomial part in $\lambda$ of the expression. Notice that the
Lax equation $\dot L=[L, L_0]$ describes the vector flow
\begin{equation}  \label{rot}
\dot{ x^{(l)}} =\langle x^{(l)},x^{(l)}\rangle y^{(l)}, \quad \dot{ y^{(l)}}
=- \langle y^{(l)},y^{(l)}\rangle x^{(l)}, \qquad
l=1,\dots, k/2 .
\end{equation}
For each index ${\cal R}_l$, equations  (\ref{rot})
generate rotations ${\cal R}_l$ in 2-planes spanned by the vectors $x^{(l)},y^{(l)}$, 
which leave the momentum $M$ invariant.

\medskip
Let $\overline{\{\cdot, \cdot \}}$ be the Poisson bracket on $\bar{\cal V}(k,m)$ obtained as
the Dirac restriction of the standard bracket in  ${\mathbb R}^{km}$. 
Symplectic properties of our system are descibed by 

\begin{proposition} \label{STAND}
The dynamical system $(\ref{r588})$ on $\bar{\cal V}(k,m)$  is Hamiltonian 
with respect to $\overline{\{\cdot, \cdot \}}$ with the Hamilton function 
$\bar H ({\cal X}, {\cal Y}) =-\frac 14 {\rm tr}(M^2 ({\cal X}, {\cal Y}) A)$. 
In the abundant coordinates ${\cal X}, {\cal Y}$ it admits the canonical representation
\begin{gather} \label{canonical}
\dot x_i ^{(l)}  = \frac {\partial \bar H}{\partial  y_i ^{(l)} } \bigg |_{\bar{\cal V}(k,m)}\, , \quad
\dot y_i ^{(l)}  =  -\frac {\partial \bar H}{\partial x_i^{(l)} }  \bigg |_{\bar{\cal V}(k,m)}\, ,  \\ 
i=1,\dots, n, \quad  l=1,\dots,k/2. \nonumber
\end{gather}
\end{proposition}

\noindent{\it Proof.\/} The equivalence of equations (\ref{canonical}) and  (\ref{r588}) on 
 $\bar{\cal V}(k,m)$ is verified by direct calculations.  
Next, according to the Dirac formalizm, the standard bracket and
$\overline{\{\cdot, \cdot \}}$ are different by terms containing $\{ f_s, \bar H \}$.
The latter equal zero since the constraint functions $f_s$ given by (\ref{constraints}) 
are invariants of the flow generated by $\bar H ({\cal X}, {\cal Y})$  on  ${\mathbb R}^{km}$. 
Hence, equations (\ref{r588}) or (\ref{canonical})  are Hamiltonian 
with respect to $\overline{\{\cdot, \cdot \}}$. $\boxed{}$
\medskip

Rotations ${\cal R}_l$ given by (\ref{rot}) are generated by the Hamiltonians $H_l(x,y)$, 
the restrictions of the functions 
$\frac 12 \langle x^{(l)},x^{(l)}\rangle \langle y^{(l)},y^{(l)}\rangle$
on $\bar{\cal V}(k,m)$. 
Clearly, these functions are first integrals of the system (\ref{r588}) and moreover 
they commute with  $H$. 

Let us fix the values of the Hamiltonians by putting 
$$
\langle x^{(l)},x^{(l)}\rangle =\langle y^{(l)},y^{(l)}\rangle =h_l, \quad h_l={\rm const}\ne 0, \quad l=1,\dots,k/2.
$$
These conditions define the customary Stiefel variety ${\cal V}(k,m)$. 
Under the substitution (\ref{r5.87}), the factor variety 
${\cal V}(k,m)/ \{ {\cal R}_1,\dots, {\cal R}_{k/2}\}$ coincides with 
a rank $k$ coadjoint orbit ${\cal S}_h^{(k)}\subset so^*(m)$ of dimension $k\left(m-\frac k2\right)-k$, 
which is parameterized by the constants $h_1,\dots,h_{k/2}$. Notice that
$M^2=h^2_1+\cdots +h^2_{k/2}$. 

\begin{theorem} \begin{description} 
\item{\rm 1)} Under the map $\bar{\cal V}(k,m) \to {\cal S}_h^{(k)}$, 
the Lie--Poisson bracket on  ${\cal S}_h^{(k)}\subset so^*(m)$  is the 
push-forward of the bracket $\overline{\{\cdot, \cdot \}}$.  
\item{\rm 2)}
The Poisson (Marsden--Weinstein)  reduction of the system (\ref{r588}) obtained by fixing values of $H_l(x,y)$ 
and by factorization by ${\cal R}_l$,  $l=1,\dots,k/2$ coincides with  the restriction of the
Frahm--Manakov system with Hamiltonian \\
$H(M)=\frac 12 \sum_{i\le j}(a_i+a_j) M_{ij}^2$ onto the orbit  ${\cal S}_h^{(k)}$. 
\end{description} 
\end{theorem}

\noindent{\it Proof.\/} 1). In view of (\ref{poinsot}),  (\ref{r5.87}), 
$$
\{ M_{ij}({\cal X}, {\cal Y}),  M_{kl}({\cal X}, {\cal Y}) \} 
=\{  M_{ij},  M_{kl}\}_{so(n)} ({\cal X}, {\cal Y})  ,
$$
i.e., the canonical bracket on ${\mathbb R}^{km}$ is the pull-back of  
the  bracket $\{\cdot, \cdot\}_{so(n)}$ on ${\cal S}_h^{(k)}\subset so^*(m)$.
On the other hand, on $\bar{\cal V}(k,m)$, 
$$
\{ M_{ij}({\cal X}, {\cal Y}),  M_{kl}({\cal X}, {\cal Y}) \} 
=\overline{  \{ M_{ij}({\cal X}, {\cal Y}),  M_{kl}({\cal X}, {\cal Y}) \}  } ,
$$
since for any $i,j,s$,  $\{ M_{ij}({\cal X}, {\cal Y}),  f_s ({\cal X}, {\cal Y}) \}=0$. 
This proves item  1). 

 2). By item  1) and Proposition \ref{STAND}, 
the  Poisson reduction of system (\ref{r588})  onto ${\cal S}_h^{(k)}$ is 
described by the Lie--Poisson bracket $\{\cdot, \cdot\}_{so(n)}$ and 
the Hamiltonian 
$H(M)=\bar H ({\cal X}, {\cal Y})= \sum_{i\le j}(a_i+a_j) M_{ij}^2$,
i.e., it is the corresponding restriction of the Frahm--Manakov system. $\boxed{}$
\medskip

The reduced system on the orbit ${\cal S}_h^{(k)}$  is integrable 
and its generic invariant manifolds are tori of dimension
$\frac 12 {\rm dim\,} {\cal S}_h^{(k)}$  (see, e.g., \cite{MiFo}). 
On the other hand, the preimage of a generic point 
$M\in {\cal S}_h^{(k)}$ in ${\cal V}(k,m)$ is a
$k/2$-fold product of circles $S^1 \times \cdots\times S^1$
(in the complex case ${\mathbb C}^*\times \cdots \times {\mathbb C}^*$). 
This implies that the original system on $\bar{\cal V}(k,m)$ has generic invariant tori of dimension
$\frac 12 {\rm dim\,} {\cal S}_h^{(k)} + k/2=(m-k/2)k/2$,
i.e., a half of dimension of the symplectic manifold  $\bar{\cal V}(k,m)$.
Hence, the original system $(\ref{r588})$ is also integrable. 

To get a global view on the above manifolds,
we represent them in the following diagram, with the dimension 
indicated above, where arrows denote the corresponding relations (restrictions or factorizations).
$$
\begin{CD}
{\mathbb R}^{km}\quad @ > f_s=0 >>
\bar{\cal V}(k,m) \quad @ > |x ^{(l)}|^2=|y ^{(l)}|^2=h_l >> 
\quad {\cal V}(k,m)  \qquad @ > {\cal R} >> \quad {\cal S}_h^{(k)}
\end{CD}
$$
$$
\boxed{k\times m} \quad \quad \boxed{k\left(m-\frac k2\right)} \qquad \quad
\boxed{ k\left(m-\frac k2\right)-\frac k2} 
\qquad \boxed{k\left(m-\frac k2\right)-k}
$$
\medskip

\noindent{\bf Remark 2.3.} 
In the case of maximal rank $k$ ($k=m$ or $k=m-1$),
when ${\cal S}_h^{(k)}$ is a generic coadjoint orbit ${\cal S}_h$, the Stiefel variety 
${\cal V}(k,m)$ is isomorphic to the group $SO(m)$. Then the following commutative diagram 
holds
$$
\begin{CD}
\bar{\cal V}(k,m) @ > |x ^{(l)}|^2=|y ^{(l)}|^2=h_l >> SO(m) \\
       @  V   {\cal R} VV        @  V   {\cal R} VV \\
so^*(m) @ > H_{l,m-l}(M)=c_l   >> {\cal S}_h \, ,
\end{CD}
$$
where the values $\{c_l\}$  of nonzero Casimir functions $H_{l,m-l}(M)$
correspond to the constants $\{h_l\}$. The mapping
$\begin{CD} SO(m) @ > {\cal R}>> {\cal S}_h \end{CD}$ 
can be regarded as a multi-dimensional analog of the Hopf fibration
$\begin{CD} SO(3) @ >S^1 >> S^2\end{CD}$.
The restriction of the system (\ref{r588}) onto ${\cal V}(k,m)$ yields an integrable
flow on the group $ SO(m)$ which describes the motion of the Frahm--Manakov top in space
for the chosen angular momentum. 

For $m=3$ such a flow was considered in \cite{Koz_vortex, Koz_AMS} from the point of
view of its hydrodynamical interpretation. 
\medskip

\paragraph{A generalization of the Chasles theorem.} If the rank $k$ is not maximal, then the components of
${\cal X}, {\cal Y}$ themselves are not sufficient to form a complete set of coordinates
on  $SO(m)$ and to  determine the position of the top in space uniquely. 
However, in this case one can make use of the following geometric property described in \cite{Fe_AMS}.
 Let us fix a part of constants of motion by putting in (\ref{cal_I})
\begin{equation} \label{c's}
\widetilde {\cal I}_{k}(s,M)=c_{0}(s-c_{1})\cdots(s-c_{m-k}), \qquad 
c_{0}, c_{1}, \dots, c_{m-k}=\mbox{const} 
\end{equation}
and consider family of confocal cones in ${\mathbb R}^{m}=(X_1,\dots,X_n)$
\begin{equation}
\bar Q(s)= \biggl\{{X^{2}_{1}\over s-a_{1}} +\cdots+
{X^{2}_{n}\over s-a_{n}} =0\biggr\} .
\end{equation}
Let $\bar {\Lambda} \subset {\mathbb R}^{m}$ be a $k$-plane spanned by the orthogonal vectors 
$x^{(1)},y^{(1)},\dots ,  x^{(k/2)},y^{(k/2)}$. 

\begin{proposition}{\rm (\cite{Fe_AMS}).} \label{prop5.6} 
\begin{description}
\item{\rm 1).} Under the motion of the Frahm--Manakov top with constants (\ref{c's})
the  $k$-plane $\bar{\Lambda}$ is tangent to the fixed cones $\bar Q (c_{1}),\ldots,\bar Q(c_{m-k})$.
\item{\rm 2).} Let $\phi^{(\alpha)}$ be a normal vector of the cone $\bar Q(c_{\alpha})$ 
at a point of the contact line $\bar\Lambda \cap \bar Q(c_{\alpha})$. 
Then the vectors $\phi^{(1)}, \dots, \phi^{(m-k)}$ together with 
$x^{(1)}, y^{(1)}, \dots, x^{(k/2)}, y^{(k/2)}$ form an orthogonal frame in ${\mathbb R}^{m}$ 
which is fixed in space. 
\end{description}
\end{proposition}

For fixed polynomial $\widetilde {\cal I}_{k}(s,M)$, the vectors  $\phi^{(l)}$ can be calculated in 
terms of $x^{(s)}, y^{(s)}$ and, thereby, the position of the top in space is completely determined. 
Proposition \ref{prop5.6} defines a single-valued map  $\bar{\cal V}(k,m)\to SO(m)$ under which
generic invariant tori of dimension $(m-k/2)k/2$ on $\bar{\cal V}(k,m)$ become tori of the same
dimension on the group $SO(m)$.

Note that the above proposition generalizes the celebrated Chasles theorem on the propery of
the tangent line to a geodesic on a quadric.

\paragraph{The rank 2 case.} 
 In the simplest case $k=2$ the angular momentum can be represented in form 
\begin{equation}\label{decom}
M=x \wedge y, \quad x=x^{(1)}=(x_1,\dots,x_m)^T  \, ,\quad 
y=y^{(1)}=(y_1,\dots,y_m)^T 
\end{equation}
and equations (\ref{r588})  describe a Hamiltonian system on the extended Stiefel variety  
$\bar{\cal V}(2,m)=\left\{ (x,y) \bigg|\,|x|=|y|, \; \langle x,y\rangle =0\right\}$, 
\begin{equation}
\label{bg}
\begin{aligned}
\dot x &=-\langle y, a x\rangle x +\langle x, a x\rangle y + a y \langle x,x\rangle , \\
\dot y &=-\langle y, a y\rangle x +\langle x,a y\rangle y -a x \langle y,y\rangle  
\end{aligned}
\end{equation}
with the Hamiltonian 
$$ \bar H= \frac 12  \langle x,a x\rangle \langle y,y\rangle 
-\langle a x,y\rangle \langle x,y\rangle  
+ \frac 12  \langle y,a y\rangle   \langle x,x\rangle 
= \frac 12  \sum_{i<j}^m (a_i+a_j) M_{ij}^2 .
$$ 

Equivalently, this system describes the evolution of fixed orthogonal vectors $x,y$ in a frame attached 
to the $m$-dimensional body.
The system admits the following $2\times 2$ Lax pair arising from (\ref{r5.89}),
\begin{gather}
\dot L(\lambda)=[\, L(\lambda), {\cal A} (\lambda)\,]\, , \label{r5.99} \\
L(\lambda)=\Phi(\lambda) \begin{pmatrix}
-\sum_{i=1}^m \frac{x_i y_i}{\lambda-a_i} &
-\sum_{i=1}^m \frac{x_i^2}{\lambda-a_i} \\ \sum_{i=1}^m \frac{y_i^2}{\lambda-a_i} &
\sum_{i=1}^m \frac{x_i y_i}{\lambda-a_i} \end{pmatrix}\, , \nonumber \\
\medskip \nonumber
{\cal A} (\lambda)=\begin{pmatrix}
-\sum_{i=1}^m (\lambda+ a_i){x_i y_i}  & -\sum_{i=1}^m (\lambda + a_i)x_i^2  \\
\sum_{i=1}^m (\lambda + a_i)y_i^2 &
\sum_{i=1}^m  (\lambda + a_i) {x_i y_i} \end{pmatrix}\, ,  \nonumber \\
\Phi(\lambda)=(\lambda-a_{1})\cdots(\lambda-a_{m}),  \nonumber 
\end{gather}

The Lax representation (\ref{r5.99}) was first indicated in
\cite{AHH_Lett}, where it was shown to be dual to an $n\times n$ Lax pair 
for the rank 2 case found by Moser in \cite{Moser_Chern}.

In view of relation (\ref{decom}), the characteristic polynomial  
$|L(\lambda)-\mu {\bf I} |$ for (\ref{r5.99}) can be written in form
$\Phi(\lambda) \widetilde {\cal I}_2(\lambda,M) +\mu^2$,
where $\widetilde {\cal I}_2(\lambda,M)$ is the family of quadratic integrals defined in (\ref{cal_I}),
\begin{align}
\widetilde {\cal I}_2(\lambda,M) &
=\sum_{i<j} \frac { \Phi(\lambda)}{(\lambda-a_{i})(\lambda-a_{j})}M_{ij}^2
\nonumber \\
& =\lambda^{m-2} H_{2,m-2}(M)+\lambda^{m-3} H_{2,m-3}(M)+\cdots+H_{20}(M) .
\label{m2.13} 
\end{align}
Notice that $H_{2,m-2}(M)= \sum_{i<j}^m M_{ij}^2=(y,y)(x,x)$ is a Casimir function
of the standard Lie--Poisson bracket on $so^*(m)$. With respect to the Poisson bracket 
$\overline{\{\, , \, \} }$ on $\bar{\cal V}(2,m)$, this function  generates permanent
rotations of the top in the fixed 2-plane $\bar\Lambda ={\rm span}(y,x)$,
which leave the components of $M$ invariant. 

Let us fix the constants of motion by putting
\begin{equation}
\label{I_2}
\widetilde {\cal I}_2(\lambda,M) =c_0(\lambda-c_1)\cdots (\lambda -c_{m-2}) , \qquad 
c_0, c_1,\dots, c_{m-2}=\mbox{const}. 
\end{equation}
This defines hyperelliptic spectral curve in ${\mathbb C}^2=(\lambda,\mu)$ of genus $g=m-2$
\begin{equation} \label{hypp}
{\cal C}=\{ \mu^2= -c_0\,  \Phi(\lambda)\, (\lambda-c_1)\cdots (\lambda -c_{m-2})\} \, .
\end{equation}

As noticed in \cite{Acta_bill}, the real generic $(g+1)$-dimensional invariant 
tori of the system can be extended to open subsets of generalized Jacobian varieties 
Jac$({\cal C},\infty_{\pm})$, which are extensions of the customary $g$-dimensional 
Jacobian Jac$({\cal C})$ by ${\mathbb C}^*$ and which can be regarded as the factor of
${\mathbb C}^2$ by the lattice generated by $(2g+1)$ independent period vectors of 
$g$ holomorphic differentials $\bar \omega_1\dots,\bar\omega_g$ and a meromorphic differential
of the third kind  $\varOmega_{\infty_{\pm}}$ having a pair of simple poles at the infinite
points $\infty_{\pm}$ on the curve $\cal C$. 

The coefficients of the matrix polynomial $L(\lambda)$ are meromorphic functions on
Jac$({\cal C},\infty_{\pm})$, whereas the components of the momentum $M_{ij}$ and 
the normal vectors $\phi^{(\alpha)}$ are meromorphic on a covering of the
Jacobian Jac$({\cal C})$ itself (the ${\mathbb C}^*$-extension is factored out by
the action of ${\cal R}=SO(2)$).

In the classical case $m=3$ the curves $\cal C$ become elliptic ones and 
generic invariant tori in $\bar{\cal V}(2,3)$ and in $SO(3)$ are 2-dimensional. 
An explicit solution for the components of the rotation matrix in terms of theta-functions
and exponents was first given in \cite{Jac_corps} (see also \cite{Whitt}).
Since now rank $M=$2 in the generic case, the above commutative diagram takes the form
$$
\begin{CD}
\bar{\cal V}(2,3) @ > |x|^2=|y|^2=h >> SO(3) \\
       @  V   SO(2) VV        @  V   SO(2) VV \\
so^*(3) @ > \langle M,M\rangle =h^2   >> S^2_h 
\end{CD}
$$
$S^2_h$ being the coadjoint orbit (2-dimensional sphere) corresponding to the constant $h$.

\section{B\"acklund transformation on $\bar{\cal V}(2,3)$, $SO(3)$ \\ 
and discretization of the clasical Euler top}

A first integrable discretization of the $m$-dimensional Euler--Manakov top was 
constructed in \cite{Ves, MosVes} by the method of factorization of matrix polynomials. 
It was represented by the correspondence 
$(\Omega,M)\to (\widetilde  \Omega, \widetilde  M)$, $\Omega\in SO(m)\,, M\in so^*(m)$, which, in
our notation reads
\begin{equation} \label{step}
M=\Omega^T A- A\Omega, \quad  \widetilde  M=\Omega M\Omega^T .
\end{equation}
Given $\widetilde  M$,  the new matrix $\widetilde \Omega$ is found from equation 
$\widetilde  M=\widetilde \Omega^T A- A\widetilde \Omega$, whose solution is not unique. 

 In given section we describe a symplectic map 
$\bar{\cal B}_{\lambda*}  \; : \bar{\cal V}(2,3) \to \bar{\cal V}(2,3)$, 
$\bar{\cal B}_{\lambda*}(x,y) =(\widetilde  x,\widetilde  y)$
governed by an arbitrary parameter $\lambda^*\in {\mathbb C}$,  which preserves the first 
integrals of the continuous system (\ref{bg}) and whose restriction 
to each generic complex torus, generalized Jacobian Jac$({\cal C},\infty_{\pm})$, is given by shift by the
2-dimensional vector 
$$
S=\int_{E_-}^{E_+} 
\left(\frac{d\, \lambda}{\mu} ,\frac{\lambda\, d\, \lambda}{\mu} \right)^T, \qquad 
E_{\pm}=(\lambda^*,\pm \mu^*), 
$$
$E_{\pm}$ being involutive points on the elliptic spectral curve, 
the simplest case of (\ref{hypp}),
$$
{\cal C}= \{\mu^2= -c_0\, (\lambda-a_1)(\lambda-a_2)(\lambda-a_3)\, (\lambda-c_1) \}.
$$
Note that $S$ is a correctly defined vector in the generalized Jacobian:
under a change of integration path on $\cal C$ it increases by a period vector
of Jac$({\cal C},\infty_{\pm})$. 
We also emphasize that here $\lambda^*$ is a constant parameter, whereas 
the conjugated coordinate $\mu^*$ depends on the equation of the curve. 

\paragraph{B\"acklund transformation on $\bar{\cal V}(2,3)$.} 
As shown in \cite{Sabaudia} by applying an addition theorem for a class of meromorphic functions on
generalized hyperelliptic Jacobians, such a map admits intertwining relation (discrete Lax pair)
\begin{gather}
\widetilde  L(\lambda) M(\lambda|\lambda^*)=M(\lambda|\lambda^*) L(\lambda), \label{bak4.8} \\ 
M(\lambda|\lambda^*)=
\begin{pmatrix} -\alpha (\lambda-\lambda^*)+ \beta & 1 \\
-\beta^2 &   -\alpha (\lambda-\lambda^*)-\beta  \end{pmatrix}, \nonumber \\
\begin{aligned}
\beta(\lambda^*) & = - \frac{\mu^*+L_{11}(\lambda^*)}{L_{12}(\lambda^*)}\equiv 
- \frac{\mu^*+ a_1^* a_2^* a_3^* \langle x,(a-\lambda^* {\bf I} )^{-1}y\rangle }
{a_1^* a_2^* a_3^* \langle x,(a-\lambda^* {\bf I} )^{-1}x\rangle }, \\
\alpha & =\frac{\partial\beta(\lambda)}{\partial\lambda} \bigg |_{\lambda=\lambda^*} ,
\quad a^*_i = a_i -\lambda^*, 
\end{aligned}
\label{ab}
\end{gather}
where $L(\lambda)$ is defined in  (\ref{r5.99}) and $\widetilde  L(\lambda)$ depends on the new variables 
$\widetilde  x, \widetilde  y$ in the same way as $L(\lambda)$ depends on $x, y$.  
In view of (\ref{m2.13}) for $m=3$, 
\begin{equation} \label{*}
\mu^*=\sqrt{\Phi (\lambda^*) \sum_{k=1}^3 a_k^* (x_i y_j- x_j y_i)^2}\, , \qquad 
(i,j,k)=(1,2,3).
\end{equation}

Now putting in (\ref{bak4.8}) subsequently  $\lambda=a_1,a_3,a_3$
and calculating the matrices $M(a_i) L(a_i|x, y) M^{-1}(a_i)$, we find
\begin{align}
{\widetilde  x}_i^2 &=\frac{ (y_i +\beta x_i+ \alpha (a_i-\lambda^*) x_i)^2} { \alpha^2 (a_i-\lambda^*)^2}, 
\nonumber \\
{\widetilde  y}_i^2 &=\frac{ (\beta y_i +\beta^2 x_i - \alpha (a_i-\lambda^*) y_i)^2 }
{ \alpha^2 (a_i-\lambda^*)^2},  \label{sqared} \\
\widetilde  x_i \widetilde  y_i &=- \frac{(y_i +\beta x_i+ \alpha (a_i-\lambda^*) x_i) \,
(\beta y_i +\beta^2 x_i- \alpha (a_i-\lambda^*) y_i)} {\alpha^2 (a_i-\lambda^*)^2}. \nonumber
\end{align}
From here the new variables can be recovered up to the action of the group generated by reflections
$(\widetilde  x_i, \widetilde  y_i)\to (-\widetilde  x_i, -\widetilde  y_i)$. Imposing the condition of the existence of 
a continuous limit (see below), we choose the following relations
\begin{gather}
\widetilde  x_i- x_i= \frac{ y_i +\beta x_i}{\alpha (a_i-\lambda^*)}, \quad
\widetilde  y_i- y_i= -\frac{\beta(y_i+\beta x_i)}{\alpha (a_i-\lambda^*)}, \qquad 
i=1,2,3 .  \label{4.10}  
\end{gather}
These expressions together with (\ref{ab}), (\ref{*}) describe the map
 $\bar{\cal B}_{\lambda*}  \; : \bar{\cal V}(2,3) \to \bar{\cal V}(2,3)$ 
in an {\it explicit\/} form. 
Since a generic parameter $\lambda^*$ corresponds to two values of $\mu^*$,
the map is generally two-valued. 
\medskip

\paragraph{Geometric model.} 
The restriction of the map onto the group $SO(3)$ admits a transparent geometric interpretation,
which can be regarded as a ``discrete version'' of the kinematic Poinsot model (see, e.g., \cite{Whitt}).
Namely, let $|x|=|y|=1$ and let 
$$R=|| x \; y \;\, x\wedge  y || \in SO(3)$$ 
be rotation matrix defining a position of a rigid body 
in space. We attach to the body a cone 
$K_2=\{(X,(a-\lambda^* {\bf I} )^{-1}X)=0\}$, which is fixed in the body
frame $(X_1,X_2,X_3)$, and assume that 
\begin{equation}
\label{real}
0< a_1 < a_2 <a_3,  \quad a_1 < \lambda^* < a_2 \quad \mbox{or} \quad a_2 < \lambda^* < a_3 .
\end{equation}
Under these conditions the cone is real and regular. 
Let $\Pi$ be 2-plane spanned by $x,y$, which is thus fixed in space and orthogonal 
to the momentum vector $M=x\wedge  y$. Assume also that $x,y$ are such that  $\Pi$ has 
a nonempty real intersection with the cone $K_2$ along lines $L_1, L_2$. 
One can show that under this condition the coordinates $\mu^*$ defined in (\ref{*})
and the parameters $\alpha, \beta$ are real. 

\begin{theorem} \label{proj_bill}
Let $K_1=\{(X,(a- h {\bf I})^{-1}X)=0\}$, $h=\textup{const}$ be a unique cone attached to the body
such that it is confocal to $K_2$  and tangent to the fixed plane $\Pi$. 
Then then new position of the body defined by the rotation matrix 
$\widetilde  R=|| \widetilde  x\;\widetilde  y\;\,\widetilde  x\wedge \widetilde y ||$ and expressions (\ref{4.10}) 
is obtained from the original position by rotating the cones
$K_1, K_2$ about axis $L_1$ or $L_2$ until $K_1$ again touches $\Pi$. 
\end{theorem}

This geometric construction is illustrated on Figure 1. 
In the new position $\widetilde  R$ determined by rotation about $L_2$, the cone $K_2$ intersects $\Pi$
along $L_2$ and another line $L_3$. Then the next iteration is generated by rotation about $L_2$ or $L_3$.
The two-valuedness of the map 
$\bar{\cal B}_{\lambda*}$  
is now related to the possibility of rotation about two different axes in ${\mathbb R}^3$. 
It follows that $N$-th iteration of the map is only $(N+1)$-valued, not $2^N$-valued. 
By fixing a sign of $\mu^*$ in (\ref{*}),  $\bar{\cal B}_{\lambda*}$  becomes single-valued
and generates a sequence of  points on $SO(3)$. 

The geometric model was first proposed in \cite{Sabaudia} as a certain limit of 
a kinematical model of motion of 4-dimensional Frahm--Manakov top in space. 
\medskip

\epsfysize=7cm
\epsfxsize=13cm
\epsfbox{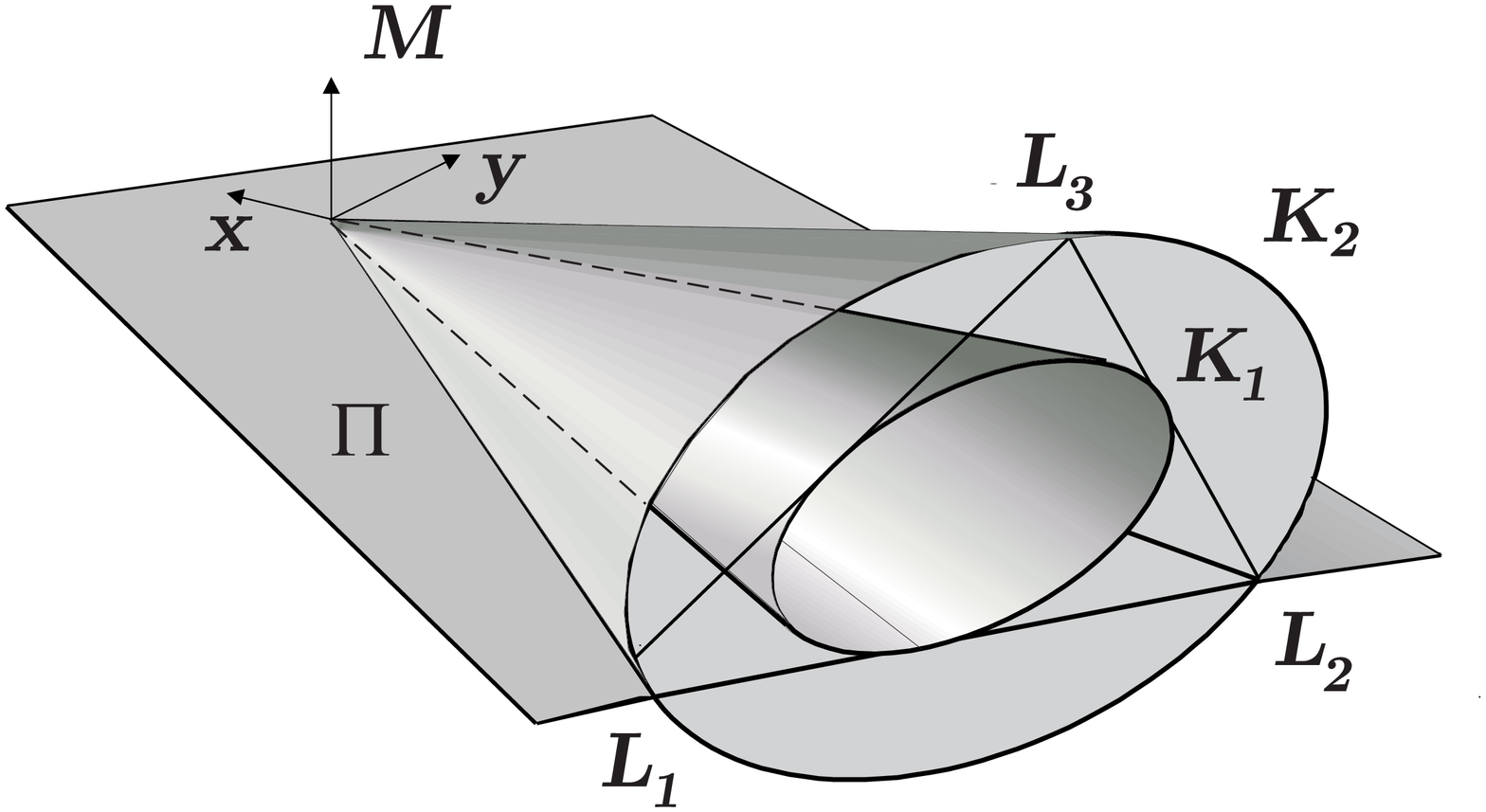}
\centerline {Figure 1}
\medskip

\noindent{\bf Remark 3.1.} As follows from (\ref{4.10}), the map $\bar{\cal B}_{\lambda*}$
admits particular solutions, for which the vector $M=x\wedge  y$ remains to be
an eigenvector of the inertia tensor $A=\mbox{diag}(a_1,a_2,a_3)$, whereas
$x, y$ themselves rotate by a fixed angle in the plane $\Lambda$. 
Such solutions can be regarded as 
analogs of stationary permanent rotations of the classical Euler top
about its principal intertia axes.

\paragraph{Continuous limit.}
Note that when $x,y$ are chosen such that $\Pi$ is (almost) tangent to the cone $K_2$ 
($c\to \lambda^*$),
$K_1$ and $K_2$ confluent and, according to the above model,  the cone $K_1=K_2$ 
is rolling without sliding 
over the fixed plane thus giving a continuous limit motion on the group $SO(3)$. 

From the algebraic geometrical point of view, in the above limit the points $E_-, E_+$
on the spectral curve $\cal C$ come together to a branch point $E_0=(c,0)$ 
and the shift vector $S$ on the generalized Jacobian tends to zero.
Let $\epsilon$ be a small complex parameter. 
Setting $\mu^*=\epsilon$, $\lambda-\lambda^*=\mbox{const}\cdot \epsilon^2$ 
in (\ref{ab}), we have the expansions
$$
\beta=-\frac {\langle x,a^{-1}y\rangle }{\langle x,a^{-1}x\rangle}+O(\epsilon), \quad 
\alpha= \frac 1\epsilon {\varkappa}{\langle x,a^{-1}x\rangle } +O(1), \qquad 
\varkappa=\frac{\partial \mu^2(\lambda)}{\partial \lambda}\bigg |_{\lambda=\lambda^*},
$$
where $\varkappa$ is a real nonzero constant. Now we set 
$$
\widetilde  x=x+\dot x\epsilon+ O(\epsilon^2), \quad 
\widetilde  y=y+\dot y\epsilon+ O(\epsilon^2) .
$$
Substituting the above expansions into (\ref{4.10}), then matching the coefficients at $\epsilon$ 
in both sides and taking into account relation 
$$ 
\langle x, (a^*)^{-1}y\rangle ^2= \langle x, (a^*)^{-1}x\rangle 
 \langle y, (a^*)^{-1}y\rangle + O(\epsilon^2),
$$
we obtain the following differential equations describing  the limit flow on a subset
of $\bar{\cal V}(2,3)$
\begin{gather} \label{cont}
\begin{aligned}
\dot x & = \frac {\det a^*}\varkappa \bigg[\langle x, (a^*)^{-1}y\rangle  (a^*)^{-1}x 
- \langle x, (a^*)^{-1}x\rangle  (a^*)^{-1}y \bigg]    
\equiv \frac 1\varkappa  x\wedge  a^*(x\wedge  y), \\
\dot y &= \frac {\det a^*}\varkappa \bigg[\langle y, (a^*)^{-1}y\rangle (a^*)^{-1}x 
- \langle x, (a^*)^{-1}y\rangle (a^*)^{-1}x \bigg]    
\equiv \frac 1\varkappa y\wedge  a^*(x\wedge  y) , 
\end{aligned} \\
a^*=a-\lambda^* {\bf I} . \nonumber 
\end{gather}
These equations are Hamiltonian with the Hamilton function 
$$
H=\frac {\det a^*}{2\varkappa} \bigg[ \langle x, (a^*)^{-1}x\rangle 
\langle y, (a^*)^{-1}y\rangle -\langle x,(a^*)^{-1}y\rangle ^2\bigg]
\equiv  \frac 1{2\varkappa} \sum_{i=1}^3 (a_i-\lambda^*) M_i^2.
$$
Notice that this function equals zero on the limit continuous flow.
The restriction of this flow on Jac$({\cal C},\infty_{\pm})$ is tangent to 
the curve ${\cal C} \subset \mbox{Jac} ({\cal C},\infty_{\pm})$ at the point $E_0$.

The above asymptotic of $\alpha, \beta$ explains the specific choice of sign of $\widetilde  x, \widetilde  y$ made in the
passage from relations (\ref{sqared}) to the map (\ref{4.10}).
\medskip

\noindent{\it Proof of Theorem} \ref{proj_bill}. 
The condition for $\Pi$ to be tangent to the cone \\ $K_1=\{(X,(a- h {\bf I})^{-1}X)=0\}$ has the form 
$$
\sum_{k=1}^3 (h-a_k) (x_i y_j- x_j y_i)^2 \equiv {\cal I}_2(h, M)=0 .
$$
Comparing this with the family of integral (\ref{I_2}) for $m=3$, we conclude that
$h=c_1$, which is constant under the map, hence the plane spanned by $\widetilde  x, \widetilde  y$ is
again tangent to $K_1$. 

Next, any translation in $SO(3)$ is represented as a finite rotation about an axis in ${\mathbb R}^3$. 
As follows from relations (\ref{4.10}),  $\widetilde  y +\beta\widetilde  x = y +\beta x$, hence 
the line along the vector  $\ell=y +\beta x$ is invariant of the action of $\bar{\cal B}_{\lambda*}$  
on ${\mathbb R}^3$ and therefore represents the axis of such a rotation. 
Finally,  in view of (\ref{ab}), we have
\begin{align*}
\langle \ell,(a^*)^{-1} \ell\rangle &= \langle y, (a^*)^{-1} y\rangle  
-2 \frac{\mu^*+L_{11}(\lambda^*)}{L_{12}(\lambda^*)}  \langle x,(a^*)^{-1}y\rangle  \\
& \quad  + \frac{(\mu^*)^2 +2\mu^* L_{11}(\lambda^*)+L_{11}^2(\lambda^*)}
{L_{12}^2(\lambda^*)} \langle x,(a^*)^{-1}x\rangle  \\
& = \frac 1 {L_{12}(\lambda^*) } 
\left[ \langle y, (a^*)^{-1} y\rangle  \langle x,(a^*)^{-1}x\rangle 
 - \langle x, (a^*)^{-1} y\rangle ^2+ (\mu^*)^2\right] ,
\end{align*}
 which equals zero by virtue of (\ref{*}).
Hence $\langle \ell,(a^*)^{-1} \ell\rangle =0$, which imply that 
the vector  $\ell$ lies on the cone $K_2$. This establishes the theorem. $\boxed{}$
\medskip

\noindent{\bf Remark 3.2.} When the attached cone $K_{2}$ does not have real intersection with 
$\Pi=\mbox{span}(x,y)$, the coordinate $\mu^*$ is imaginary and, according to (\ref{4.10}), 
(\ref{ab}), the new values $\widetilde  x, \widetilde  y$ are complex. As a result, under the reality conditions
(\ref{real}) the map $\bar{\cal B}_{\lambda*}$  is real only on the subset 
$\bar {\mathfrak R}\subset \bar{\cal V}(2,3)$
defined by unequality 
$$
\sum_{k=1}^3 (\lambda^*-a_k) (x_i y_j- x_j y_i)^2 \le 0
$$
On the boundary of $\mathfrak R$, the map  tends to the identical one. 

\paragraph{Reduction to the coalgebra $so^*(3)$.} Under the factorization by rotations of
${\cal R}=SO(2)$, the transformation $\bar{\cal B}_{\lambda*}$  induces a map 
${\cal B}_{\lambda*}\, :\, so(3)^* \to so(3)^*$  such that 
$$
\widetilde  M \equiv {\cal B}_{\lambda*} M(x,y) =\widetilde x \wedge \widetilde y. 
$$
The latter map is correctly defined, i.e., it does not depend on a concrete choice of vectors
 $x,y$ giving the same $M$. It preserves the first integrals of the 
classical Euler top on $so^*(3)$ and its generic invariant manifolds are open subsets 
of 4-fold unramified coverings of the complex torus  Jac$({\cal C})={\cal C}$. 
The restriction of ${\cal B}_{\lambda*}$ onto 
Jac$({\cal C})$ is given by shift by the holomorphic integral
$e=\int_{E_-}^{E_+} d\lambda/\mu$, which thus depends only on the constants $c_0, c_1$.
According to a theorem in \cite{Bob_Lorb},  this implies that  the map ${\cal B}_{\lambda*}$ 
preserves the standard Lie--Poisson structure on $so^*(3)$. 

\begin{proposition} \label{reflection} 
Vectors $M, \widetilde  M$ satisfy the following symmetric relations
\begin{gather} \label{b-l-s}
\widetilde  M -M = \varkappa \, (\widetilde  M + M) \wedge  a (\widetilde  M + M) , \\
\varkappa = \sqrt{2 \langle  aM+ a\widetilde  M, aM+ a\widetilde  M\rangle  }\, 
\frac {\sqrt {1- \langle M,\widetilde M\rangle/c_0}}
{\sqrt {1+\langle M,\widetilde M\rangle /c_0} }\, ,
\label{kappa} \\
 \langle M, a^* M\rangle =\langle \widetilde  M, a^* \widetilde  M\rangle 
= - \langle M,a^*\widetilde M\rangle , \label{bill}
\end{gather}
where, as above, $c_0=\langle M,M\rangle =\langle \widetilde  M,\widetilde M\rangle$.
\end{proposition}

Relation (\ref{b-l-s}) was previously obtained by  another method  in \cite{Bob_Lorb}, 
as an implicit map describing a Poisson discretization of the Euler top in $so^*(3)$. 
\medskip

\noindent{\it Proof of Proposition} \ref{reflection}. 
In view of relations  (\ref{4.10}),  we find
\begin{equation} \label{times}
\widetilde x \wedge \widetilde y =\widetilde M
= x\wedge  y+ \frac 1\alpha \left(a^*\right)^{-1} \ell \wedge \ell, 
\end{equation}
where, as  above,  $\ell=y+\beta x$,  $a^*=a-\lambda^*{\bf I}.$
Note that vector $\left(a^*\right)^{-1} \ell$ is normal to the cone $K_2$ at a point of the intersection
line $L_2$ or  $L_1$.  Hence, $\widetilde  M -M$ is orthogonal to $\left(a^*\right)^{-1}\ell$ and $\ell$.
Next, since $\ell$ lies in the planes $\Pi, \widetilde  \Pi$, this vector is orthogonal to $M, \widetilde  M$.
This, together with the equality $|M|=|\widetilde  M|$ implies that the sum $\widetilde  M +M$ is
parallel to $\left(a^*\right)^{-1} \ell$ and  $a^*(\widetilde  M +M)$ is parallel to $\ell$. 
As a result, (\ref{times}) implies (\ref{b-l-s}).

To find factor $\varkappa$, we first introduce angle $\phi$ between vectors 
$\widetilde  M$ and $M$. Since $|M|=|\widetilde  M|$, the vectors $\widetilde  M -M$ and $\widetilde  M+M$ 
are orthogonal, and we have
$$
|\widetilde  M -M|= \frac 12 |\widetilde  M+M|\tan \frac\phi 2 
\equiv \frac 12|\widetilde  M+M| \frac {\sqrt {1-\langle M,\widetilde M\rangle/c_0 }}    
{\sqrt {1+\langle M,\widetilde M\rangle/c_0}} \, .   
$$
On the other hand, since $a^*(\widetilde  M+M)$ is orthogonal to $\widetilde  M, M$, from 
(\ref{b-l-s}) and the properties of the vector product we deduce 
$$
|\widetilde  M -M|= \varkappa \,|\widetilde  M+M|\, |a^*(\widetilde  M+M)|. 
$$
Comparing the right hand sides of the above two relations, we obtain (\ref{kappa}).

The first equality in (\ref{bill}) holds because the map  ${\cal B}_{\lambda*}$ preserves the first integrals
of the Euler top. Next, since the vector $a^*(\widetilde  M +M)$ lies on the cone
$K_2$, we have $((\widetilde  M +M), a^*(\widetilde  M +M))=0$. Expanding this and using the
 first equality in (\ref{bill})  yields the second equality. $\boxed{}$
\medskip

\noindent{\bf Remark 3.3.} 
 The fact that the difference $\widetilde  M -M$ is orthogonal to $\left(a^*\right)^{-1}\ell$ and $\ell$ implies 
that {\it the angle between the normal vector $\left(a^*\right)^{-1}\ell$
and the plane $\Pi$ equals the angle between $\left(a^*\right)^{-1}\ell$ and \/} $\widetilde \Pi$. 
This property can be regarded as a projective version of the Birkhoff condition of elastic impacts,
hence the geometric construction of Theorem \ref{proj_bill} illustrated in Figure 1 describes  a projective analog of
the plane  elliptic billiard. (Note that no any plane section of the cones and of the sequence of $\Pi$ 
gives such a plane billiard.)
 \medskip

To obtain the map ${\cal B}_{\lambda*}\, :\, so^*(3) \to so^*(3)$  in an explicit form, 
we use the fact that the vector $\ell$ satisfies the system of homogeneous equations
$$
\langle \ell,(a^*)^{-1} \ell\rangle=0,  \quad \langle \ell, M\rangle =0 . 
$$
One of its solutions, 
$ \bar\ell =(\bar\ell_1, \bar\ell_2, \bar\ell_3)^T$, normalized by the condition $\bar \ell_3=1$, 
has the form
\begin{gather}
\bar \ell_1= - a_1^* \frac {M_1 M_3 - \sqrt{D}\,  a_2^* M_2 }{a_1^* M_1^2 + a_2^* M_2^2},  \quad
\bar\ell_2= - a_2^*  \frac {M_2 M_3 + \sqrt{D} \, a_1^* M_1}{a_1^* M_1^2 + a_2^* M_2^2}, \quad
\bar \ell_3=1 \\
D= \frac{a_1^* M_1^2+ a_2^* M_2^2 +a_3^* M_3^2} {a_1^* a_2^* a_3^*} . \label{D}
\end{gather}
Substituting this instead of $\ell$ into (\ref{times}) and symmeterising obtained 
expressions, we arrive at the following relations
\begin{equation} \label{diff}
\begin{aligned}
\widetilde  M_1 -M_1 &=\chi (a_2 - a_3) 
\Big( \Delta_1\,  M_2 M_3 
  - \sqrt{D}\, (a^*_{1}) ^{2} M_1^3 ( a_2^* M_2^2 -a_3^* M_3^{2}) \Big) , \\
\widetilde  M_2 -M_2 &=\chi (a_3 - a_1) \Big( \Delta_2\, M_1 M_3 
- \sqrt{D}\, (a_{2}^*)^{2} M_2^{3} ( a_3^* M_3^2 -a_1^* M_1^2 ) \Big) , \\
\widetilde  M_3 -M_3 &=\chi (a_1 - a_2) 
\Big(\Delta_3 M_1 M_2 - \sqrt{D}\, (a_{3}^*)^2 M_3^{3}(a_1^* M_1^2 -a_2^* M_2^2) \Big) ,
\end{aligned}
\end{equation}
where 
$$
\Delta_i=a_{j}^* a_{k}^* M_j^2 M_k^2 - (a^*_i)^2 M_i^4 , \qquad 
(i,j,k)=(1,2,3)
$$
and $\chi$ is a common factor.

Next, multiplying the both sides of (\ref{diff}) by $a_1^*M_1, a_2^*M_2, a_3^*M_3$ 
respectively, then summing and using the second equality in (\ref{bill}),  
we find the factor $\chi$ in form
\begin{equation} \label{chi}
\chi= - \frac{2\langle M, a^*M\rangle} {\sum a_1 (a_2 - a_3) \bigg[ 
\Delta_1 M_1 M_2 M_3 
-\sqrt{D}(a^*_{1})^{2} M_1^4 ( a_2^* M_2^2 -a_3^* M_3^{2})\bigg]}\, ,
\end{equation}
where the summation in the denominator ranges over the three terms obtained by 
the cyclic permutations of indices (1,2,3). Thus the right hand sides of (\ref{diff})
are homogeneous expressions of degree -1 in $M_i$.
\medskip

We summarize our results on the map ${\cal B}_{\lambda*}$ in the following theorem.  

\begin{theorem} The map ${\cal B}_{\lambda*}\, :\, so^*(3) \to so^*(3)$  given by
(\ref{diff}), (\ref{chi}), (\ref{D}) preserves the first integrals of the continuous Euler top,
as well as the standard Lie--Poisson structure on $so^*(3)$.
Under reality conditions (\ref{real}), the map is real inside the conical domain
$$
{\mathfrak R} =\left\{ \sum_{k=1}^3 (\lambda^*-a_k) M_k^2 \le 0\right \}.
$$
Its restriction onto each regular elliptic curve ${\cal C}$ is represented by the shift by 
the holomorphic integral $e=\int_{E_-}^{E_+} d\lambda/\mu$. Near the boundary of 
${\mathfrak R}$ the energy constant $c_1$ tends to $\lambda^*$ and 
the shift $e$ tends to zero. 
The continuous limit of ${\cal B}_{\lambda*}$ coincides with the Euler equations
$\dot M=[M,aM]$.
\end{theorem}

\paragraph{Remark 3.4.}
Since generic solutions $M_1(t), M_2(t), M_3(t)$ of the classical Euler top are 
proportional to the elliptic functions
$\textup{sn}(u), \textup{cn} (u), \textup {dn}(u)$, with $u=\mbox{const}\cdot t$ 
and the map ${\cal B}_{\lambda*}$ results in adding $e$ to the argument $u$, 
the relations (\ref{diff}), (\ref{D}), (\ref{chi}) can be regarded as
a set of explicit addition formulae for these functions.
However, in this case neither the moduli of the curve $\cal C$ nor the shift parameter
are fixed: they depend on initial position of $M$ in $so^*(3)$.
 

\end{document}